\documentclass{article}
 
\usepackage{amssymb,amsmath,amsthm} 
\usepackage{graphicx}
\usepackage[latin1]{inputenc}
\usepackage[a4paper, centering, text={6.5in, 10in}]{geometry} 
\usepackage{authblk}

\newtheorem{example}{Example}[section]

\providecommand{\keywords}[1]{\textbf{\textit{Index terms---}} #1}

\author{Francisco Mota}
\affil{Departamento de Engenharia de Computa\c c\~ao e Automa\c c\~ao\\
Universidade Federal do Rio Grande do Norte -- Brasil\\
e-mail:mota@dca.ufrn.br}

\date{\today}

\title{Splitting Root-Locus Plot into Algebraic Plane Curves}

\begin{document}

\maketitle

\begin{abstract}
In this paper we show how to split the root-locus plot for an irreducible rational transfer function
into several individual algebraic plane curves, like lines, circles, conics, etc. To achieve this goal we
use results of a previous paper of the author to represent the Root Locus as an algebraic 
variety generated by an ideal
over a polynomial ring, and whose primary decomposion allow us to isolate the planes curves that 
composes the Root Locus. As a by-product, using the concept of duality in projective algebraic geometry,
we show how to obtain the dual curve of each plane curve that composes the Root Locus and 
unite them to obtain what we denominate the ``Algebraic Dual Root Locus''.

\keywords{Root-Locus, Projective Root-Locus, Ideal of Polynomials, Primary Decomposition, 
Real Projective Plane, Dual Algebraic Curve, Grobner Basis.}

\end{abstract}

\section{Introduction}

\begin{figure}
\begin{center}
\includegraphics{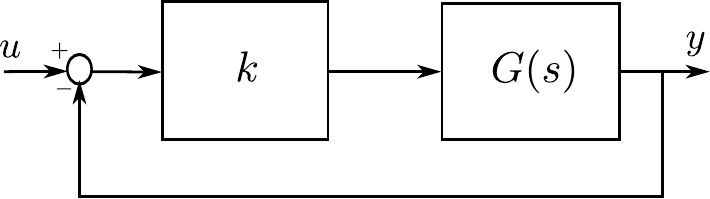}
\caption{\label{cloop}Control Feedback Loop with a Proportional Controller}
\end{center}
\end{figure}

Root-Locus (RL) is a  parametric plot of the roots of the polynomial
$p(s) = d(s) + kn(s)$  over the complex plane, equivalently over the affine plane $\mathbb{R}^2$,
as the parameter $k$ spans $\mathbb{R}$; $d$ and $n$
are fixed coprime polynomials, and $d$ is monic with degree, in general, greater
than the degree of $n$. The polynomial $p$ can represent the denominator of a (proportional) control
feedback loop of a  linear time invariant plant  with transfer function $G(s)=n(s)/d(s)$ (see Figure~\ref{cloop}), 
and this makes  the RL a classical approach to study stability and performance of 
closed loop feedback systems. The rules for sketching the plot are discussed in most textbooks on 
feedback control theory of linear systems (see \cite{dh}).
In a previous paper (\cite{pjrl}) the author showed that the RL plot can be extended 
to the  real projective plane ($\mathbb{RP}^2$) and be interpreted as a projective algebraic 
variety, that we denominated projective root-locus (PjRL). 
In this approach, the RL points, including the ones at infinity, 
can be calculated as the solution of a set of homogeneous polynomial equations, 
as well as, we can obtain complementary 
plots of RL over the different affine planes that make up the projective plane.

In this paper we use the approach presented in \cite{pjrl} to show how to decompose the RL into several
planes curves that can be plotted independently to form the final RL plot. In fact, at least in some simple cases, 
we can easily visualize the RL as a union of several plane curves: for example, 
the plot presented in Figure~\ref{rlc}, that represents the RL for $G(s)=(s+1)/s^2$, 
is composed by the (parametrized) circle $(x+1)^2+y^2=1$ and by the (parametrized) line $y=0$. In order to deal with this 
question in a systematic way, however, we need concepts from algebraic geometry, considering
the RL as an algebraic variety, and the goal is to find its decomposition into {\em irreducible components} 
(see \cite[Chap. 4]{clo}).
We cannot, in general, obtain the irreducibles components of an algebraic variety ``by hand", 
but considering the curve as the set of zeros of an ideal in a polynomial ring, the question
becomes strongly related to the computation of {\em primary decomposition} of ideals (see \cite[Chaps. 4,7]{am}), 
a fundamental topic in abstract algebra, and for which computing algorithms there exists
since a long time (\cite{wiki}). In particular, 
Macaulay2 package (\cite{mac}) incorporates a command to
compute the primary decompositon of an ideal.

We also present in this paper, mainly as a matter of mathematical curiosity, a new root-locus plot that we denominate
``Algebraic Dual Root-Locus'' or (ADRL), associated to the conventional RL plot, that is 
obtained by computing the dual curve (in projective geometry sense) 
for each individual plane curve that makes up the 
projective root-locus. We leave the analysis of the properties of ADRL for a possible future work. 

Bellow we present some concepts used in the paper: 

\begin{figure}
\begin{center}
\includegraphics[scale=1.5]{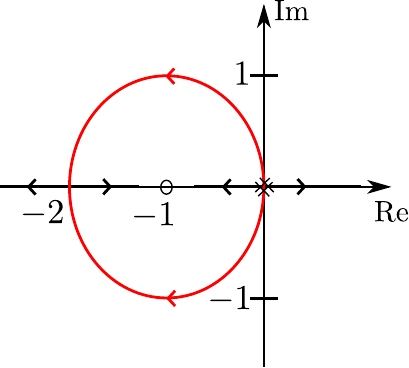} 
\caption{\label{rlc} Root-Locus for $G(s)=(s+1)/s^2$} 
\end{center}
\end{figure} 

\begin{description}

\item[$\pmb{\mathbb R}$, $\pmb{\mathbb C}$] {\bf and} $\pmb{\mathbb R[x_1, x_2, \ldots, x_n]}$:
Represents the field of real numbers, the
field of complex numbers and the ring of polynomials with coefficient's in
$\mathbb R$ and with indeterminates 
$(x_1, x_2, \ldots, x_n)$, respectively.

\item [Homogeneous polynomial] A polynomial (in several variables) is
homogeneous when all of its nonzero terms (monomials)
have the same total degree (sum of the degree of each variable). We always
can turn a non-homogeneous polynomial ($q$) into a homogeneous one ($q^h$)
by adding a new variable ($x_{n+1}$), with the following procedure:
$q^h(x_1, \ldots, x_n,x_{n+1}) = x_{n+1}^{d}\;q(x_1/x_{n+1}, x_2/x_{n+1}, 
\ldots, x_n/x_{n+1})$, where $d$ is the total degree of $q$; this process is
denominated ``homogenization'' of $q$.   We can always ``de-homogenize"
$q^h$ by setting $x_{n+1}=1$ and recover back $q$.

\item[Ideal of Polynomials:] A set of polynomials 
$I \subseteq  \mathbb R[x_1, x_2, \ldots, x_n]$ is an ideal when
it satisfies the following properties (\cite{clo}, \cite{am}): 
$(i)$ $0 \in I$; 
$(ii)$ $p,q \in I$ implies $p+q \in I$; and 
$(iii)$ $p\in I$ and $q\in \mathbb R[x_1,x_2,\ldots, x_n]$ implies $pq\in I$.
One important fact about ideals of the ring $\mathbb R[x_1,x_2,\ldots,x_n]$ is that they
are {\em finitely generated}, that is, for every ideal $I$ always there exists a {\em finite} subset of 
polynomials in $I$, denoted by $\{p_1,p_2,\ldots, p_t\}$, such that 
\[
I = \sum_{i=1}^{t}h_ip_i, \quad h_i \in \mathbb R[x_1, x_2, \ldots, x_n].
\] 
The set $\{p_1, p_2, \ldots, p_t\}$ is denominated a {\em generating set for $I$}; in this case
we write $I=\langle p_1, p_2, \ldots, p_t\rangle$. A Grobner Basis for the ideal 
$I$ is a particular kind of generating set that allows many important 
properties of the ideal to be deduced easily. Given a generating 
set $\{p_1, p_2, \ldots, p_t\}$ for $I$, we can obtain a Grobner basis
$\{g_1, g_2, \ldots, g_s\}$ for $I$ algorithmically (see \cite[Ch.~2]{clo}).
The most basic ideals are $\langle 0\rangle$, the zero ideal, and
$\langle 1\rangle$, the ring $\mathbb R[x_1,x_2,\ldots,x_n]$ itself. In fact, 
if $1\in I$, we immediatelly conclude $I= \mathbb R[x_1,x_2,\ldots,x_n]$.
We also have that the intersection of any family of ideals results in an ideal.
Another important result related to ideals in Noetherian rings (like
$\mathbb R[x_1,x_2,\ldots,x_n]$) is the ``Lask-Noether Theorem" which 
states that every ideal in a Noetherian ring can be written as an finite intersection
of primary ideals.

\item[Variety generated by an ideal:] An (real) algebraic variety is a subset of $\mathbb R^n$ 
whose elements are the (real) solutions a system of polynomial equations in $n$ variables
 (in $\mathbb R[(x_1,x_2,\ldots,x_n])$.
We can see this set of polynomials as a generating set of an ideal $I$, and so we say
that the variety is generated by the ideal $I$, and represented by $\mathbb V(I)$. 
There exists several important relationships between ideals and verieties, in particular: 
$\mathbb V(\langle 0\rangle)=\mathbb R^n$, $\mathbb V(\langle 1\rangle)=\emptyset$ and 
if $I$ and $J$ are ideals we have that $\mathbb V(I\cap J) = \mathbb V(I) \cup \mathbb V(J)$.

\end{description}
For more details about the concepts above see (\cite{clo}, \cite{am}).

\section{Decomposing projective root-locus into irreducible components}\label{ic}

In a previous paper (\cite{pjrl}) the author showed how to extended the RL plot 
from the affine plane $\mathbb R^2$ to the projective plane $\mathbb{PR}^2$
by considering the parametric plot roots of the ``modified" polynomial 
\begin{equation}\label{mpoly}
p(s) = k_dd(s)+k_nn(s)
\end{equation}
over $\mathbb{PR}^2$ as $k_n/k_d$ spans the 
projective line $\mathbb{PR}^1$. Considering the ideal $I=\langle u, v \rangle$, 
generated by the polynomials 
$u=\text{Re}\{p(x+iy)\}$ and $v=\text{Im}\{p(x+iy)\}$, 
the projective root-locus (PjRL) is obtained from the
Grobner basis $\{g_{1}, g_{2},\ldots, g_{s}\}$ for the ideal $I$, 
defined in the ring $\mathbb R[x,y,k_d,k_n]$, 
with respect a graded monomial order. If we define the homogenization
of the ideal $I=\langle u, v\rangle$ as the ideal 
$I^h=\langle g_{1}^h, g_{2}^h, \ldots, g_{s}^h\rangle$, where $g_{i}^h$ is the (homogeneous)
polynomial obtained by the homogenization of $g_{i}$, we can obtain the projective root-locus
from the (projective) variety generated by the ideal $I^h$, denoted by $\mathbb V(I^h) $,
where $I^h $ is defined in the ring 
$\mathbb R[x,y,z,k_d,k_n]$ (see \cite{pjrl} for details).

To decompose the PjRL into irreducible components we need to obtain 
a primary decomposition of $I^h$, in order to write it  as a
finite intersection of ideals, that is
\begin{equation}\label{pdec}
I^h = J_1 \cap J_2 \cap \cdots \cap  J_m,
\end{equation}
where each $J_i$ is a primary ideal (see \cite[Thm. 7.13]{am}).
Based on this, we can write $\mathbb V(I^h) $, the variety generated
by $I^h $, as 
\[
\mathbb V(I^h) = \mathbb V(J_1) \cup \mathbb V(J_2) \cup \cdots \cup \mathbb V(J_m),
\]
where $\mathbb V(J_i)$ is the variety generated by the primary ideal $J_i $, and it is
an irreducible component of the variety $\mathbb V (I^h) $. We note that, given a generating set for the
ideal $I^h $, namely the Grobner basis $\{g_{1}^{h}, g_{2}^{h}, \ldots, g_{s}^{h}\} $, we can 
obtain a generating set for each primary ideal $J_i $ in (\ref{pdec}) by a computational algorithm,
like the command ``primaryDecomposition" in Macaulay2 package. In this way, we have the
following procedure for finding the irreducible components of the PjRL for an irreducible 
transfer function $G (s)=n (s)/d (s) $:
\begin{enumerate}
\item Define $p(s) = k_dd(s) + k_nn(s) $ and taking $s = x + jy $, obtain $p(x+jy) = u(x,y,k_d,k_n) + jv (x,y,k_d,k_n) $;
\item Obtain a Grobner basis for the ideal $I=\langle u, v\rangle $, and denote it by $\{g_1, g_2, \ldots, g_s\} $;
\item Let $I^h = \langle g_{1}^{h}, g_{2}^{h}, \ldots, g_{s}^{h}\rangle $, the homogenization of $I$, and obtain the 
primary decomposition of the ideal $ I^h $, as presented in Equation~(\ref{pdec});
\item The zeros of the generating set for each $J_i $ in (\ref{pdec}) is an irreducible variety, 
whose union for $i = 1, 2, \ldots, m $ makes up the PjRL.

\end{enumerate}

\subsection{Examples}

In all examples below we used the Macalay2 software (\cite{mac}) to make the calculations 
and all polinomials are defined with coeficients in the field of rationals (that is in 
$\mathbb Q[x_1,x_2,\ldots,x_n]$) so that we can get infinite precision in calculations.

\begin{example}\label{ex1}\em
Let be $G(s) = (s+1)/s^2$, whose RL plot is shown in Figure~\ref{rlc}, and
\[
p(s) = k_ds^2 + k_n(s+1).
\]
Defining $u={\rm Re}\{p(x+jy)\}$ and $v={\rm Im}\{p(x+jy)\}$ we
have:
\[
u(x,y,k_d,k_n) = k_d (x^2-y^2)+k_n(x+1), \quad
v(x,y,k_d,k_n) = 2k_d xy + k_n y.
\]
Now we compute the Grobner basis for the ideal $\langle u,v \rangle$
using the graded reversed lexicographic order with 
$x>y>k_d>k_n$ and obtain $\{g_1,g_2,g_3,g_4\}$, where:
\begin{eqnarray*}
g_1(x,y,k_d,k_n) & = & 2xyk_d+yk_n \quad (= r) \\
g_2(x,y,k_d,k_n)  & = & x^2k_d-y^2k_d+xk_n+k_n \quad (= q)\\
g_3(x,y,k_d,k_n)  & = & x^2yk_n+y^3k_n+2xyk_n \\
g_4(x,y,k_d,k_n)  & =& 2y^3k_d-xyk_n-2yk_n
\end{eqnarray*}
Homogenizing of the polynomials $g_i$, using the
procedure indicated in the Introduction we obtain:
\begin{eqnarray*}
g_{1}^{h} & = & z^3g_1(x/z,y/z,k_d/z,k_n/z) = 2xyk_d + yzk_n \\
g_{2}^{h} & = & z^3g_2(x/z,y/z,k_d/z,k_n/z) = x^2k_d-y^2k_d+xzk_n+z^2k_n \\
g_{3}^{h} & = & z^4g_3(x/z,y/z,k_d/z,k_n/z) = x^2yk_n + y^3k_n + 2xyzk_n \\
g_{4}^{h} & = & z^4g_4(x/z,y/z,k_d/z,k_n/z) = 2y^3k_d - xyzk_n - 2yz^2k_n 
\end{eqnarray*}
Now we compute the primary decomposition for the ideal 
$I^h=\langle g_{1}^{h}, g_{2}^{h}, g_{3}^{h}, g_{4}^{h}\rangle$ to
obtain $I^h = J_1 \cap J_2 \cap J_3$, where:
\begin{eqnarray}
J_1 & = & \langle y, x^2k_d + xzk_n + z^2k_n \label{j1ex1}\rangle \\
J_2 & = & \langle x^2 + y^2 + 2xz, 2xk_d +zk_n\label{j2ex1}\rangle \\
J_3 & = & \langle k_d, k_n\rangle \label{j3ex1}
\end{eqnarray}
Using the fact that $k_d$ and $k_n$ belong to the set of reals and that they can't be both
simultaneously zero, we have that $1\in J_3$ (suppose $k_n\neq 0$, so $(1/k_n)\times k_n=1\in J_3$)
and then $J_3=\mathbb R[x,y,z,k_d,k_n]$ can be deleted from the primary decomposition of
$I^h$, that is, $I^h=J_1\cap J_2$. Then we have that $\mathbb V(I^h) = \mathbb V(J_1)\cup \mathbb V(J_2)$,
where $\mathbb V(J_1)$ is defined by $y=0$ and $\mathbb V(J_2)$ is defined by $x^2+y^2 + 2xz=0$. To analyze
these varieties in the affine $XY$ plane we set $z=1$ and we obtain the components of the 
plot shown in Figure~\ref{rlc}, that is the line $y=0$ and the circle $(x+1)^2 +y^2 = 1$, as desired. Also, from 
the ideals $J_1$ and $J_2$ above we can obtain the parametrization of these curves as well as the initial and terminal
points of the PjRL, as was done by the author in \cite{pjrl}:
\begin{description}
\item[Variety $\mathbb V(J_1)$:] Defined by the ideal $J_1$, as shown in Equation~(\ref{j1ex1})
\begin{itemize}
\item Initial Points: $k_d=1$ and $k_n=0$. From (\ref{j1ex1}), we get $y=0$ and $x^2=0$ or $x=0$. Therefore the initial 
point 
for $\mathbb V(J_1)$ is $(0:0:1)$ or $(0,0)$ in affine plane $XY$.
\item Terminal points: $k_d=0$ and $k_n=1$. We get from (\ref{j1ex1}), $y=0$ and $xz+z^2=0$. Then we have (a) $z=0$ and 
$x=1$, which is
the point at infinity $(1:0:0)$ (horizontal lines) and (b) $z=1$ which implies $x=-1$ and the point is $(-1:0:1)$ or 
$(-1,0)$ in affine plane $XY$.
\item Intermediary Points: $k_d=1$ and $k_n=\lambda\neq 0$. Again from (\ref{j1ex1}), we have $y=0$ and 
$x^2+xz\lambda + z^2\lambda=0$,
and we must have $z=1$ ($z=0$ would imply $x=0$ what is impossible); so, all intermediary points are at finite position 
and is given by
$y=0$ and $x^2 + x\lambda + \lambda=0$, $\lambda\neq 0$. Then $x=\frac{-\lambda\pm\sqrt{\lambda^2-4\lambda}}{2}$, which 
give us the 
RL over $y=0$ line.

\end{itemize}

\item[Variety $\mathbb V(J_2)$:] Defined by the ideal $J_2$, as shown in Equation~(\ref{j2ex1})
\begin{itemize}
\item Initial Points: $k_d=1$ and $k_n=0$. From (\ref{j2ex1}), we get $x^2 + y^2 + 2xz=0$ and $2x=0$ or $x=0$. Then we 
have $x^2+y^2=0$
or $y=0$. Therefore the initial point for $\mathbb V(J_2)$ is $(0:0:1)$ or $(0,0)$ in affine plane $XY$.
\item Terminal points: $k_d=0$ and $k_n=1$. We get from (\ref{j2ex1}), $x^2 + y^2 + 2xz=0$ and $z=0$. Then we get 
$x^2 + y^2=0$ what
implies $x=y=z=0$ which is not allowed, so this variety has no terminal points.
\item Intermediary Points: $k_d=1$ and $k_n=\lambda\neq 0$. Again from (\ref{j2ex1}), we have $x^2 + y^2 + 2xz=0$ and 
$2x + z\lambda=0$,
and we must have $z=1$ ($z=0$ would imply $x=y=0$ what is impossible); so, all intermediary points are at finite 
position and is given by
$x^2 + y^2 + 2x = 0$ and $2x + \lambda=0$, $\lambda\neq 0$, which is the parametrized equation of the circle as 
shown in RL plot.

\end{itemize}

\end{description}

Now, since the complete PjRL is $\mathbb V(J_1) \cup \mathbb V(J_2)$, we have:
\begin{itemize}
\item Initial points: $(0:0:1)$ (from $\mathbb V(J_1)$) plus $(0:0:1)$ (from $\mathbb V(J_2)$); so we have a duplicate 
point at 
$(0:0:1)$ or at $(0,0)$ in affine plane $XY$.
\item Terminal Points: $\{(1:0:0), (-1:0:1)\}$, only from $\mathbb V(J_1)$
\item Intermediary points: $\{(x:0:1), x=\frac{-\lambda\pm\sqrt{\lambda^2-4\lambda}}{2}, \lambda\neq 0\}$ from 
$\mathbb V(J_1)$ plus 
$\{(x:y:1), x=-\lambda/2, \lambda\neq 0$ and $x^2+y^2+2x=0$\} from $\mathbb V(J_2)$.

\end{itemize}

\end{example}

\begin{example} \em \label{ex2} We now consider a modification of Example~\ref{ex1} 
above by defining $\displaystyle G(s)=\frac{s+1}{s^2(s+4)}$, whose RL plot is shown in Figure~\ref{rlwc}. Also
in this case we see that the RL is the union of the line $y=0$ and the ``weird'' curve shown in red. In this case:
\[
p(s) = k_d(s^3+4s^2) + k_n(s+1)
\]
and we have 
\begin{eqnarray*}
u={\rm Re}\{p(x+jy)\} & = &k_d(x^3 -3xy^2 + 4x^2 - 4y^2) +k_n(x+1)\\
v={\rm Im}\{p(x+jy)\} & = &k_d(-y^3 + 3x^2y + 8xy) + k_ny.
\end{eqnarray*}
The Grobner basis for the ideal $\langle u,v\rangle$ is $\{g_1,g_2,g_3,g_4,g_5\}$, and the generating set for 
$I^h$ is $\{g_{1}^{h},g_{2}^{h},g_{3}^{h},g_{4}^{h},g_{5}^{h}\}$, where:
\begin{eqnarray*}
g_{1}^{h} & = & 3x^2yk_d - y^3k_d + 8xyzk_d + yz^2k_n\\
g_{2}^{h} & = & x^3k_d - 3xy^2k_d + 4x^2zk_d - 4y^2zk_d + xz^2k_n + z^3k_n\\
g_{3}^{h} & = & 2x^3yk_n + 2xy^3k_n + 7x^2yzk_n + 3y^3zk_n + 8xyz^2k_n \\
g_{4}^{h} & = & 24xy^3k_d + 32y^3zk_d + 32xyz^2k_d - 6xyz^2k_n - 5yz^3k_n \\
g_{5}^{h} & = & 24y^5k_d + 160y^3z^2k_d - 128xyz^3k_d - 18x^2yz^2k_n - 24y^3z^2k_n - 39xyz^3k_n - 52yz^4k_n
\end{eqnarray*}
and computing the primary decomposition for $I^h$ we obtain $I^h = J_1\cap J_2\cap J_3$, where
\begin{eqnarray}
J_1 & = & \langle y, x^3k_d + 4x^2zk_d + xz^2k_n + z^3k_n\rangle \label{j1ex2}\\
J_2 & = & \langle 2x^3 + 2xy^2 + 7x^2z + 3y^2z + 8xz^2, 3x^2k_d - y^2k_d + 8xzk_d + z^2k_n\rangle \label{j2ex2}\\
J_3 & = & \langle k_d, k_n\rangle \label{j3ex2}
\end{eqnarray}
Once more, $J_3$ can be removed from the intersection, so $\mathbb V(I^h)=\mathbb V(J_1)\cup \mathbb V(J_2)$,
where $\mathbb V(J_1)$ is defined by $y=0$ and $\mathbb V(J_2)$ is defined by 
$2x^3 + 2xy^2 + 7x^2z + 3y^2z + 8xz^2=0$. As in Example~\ref{ex1} above, to analyze
these varieties in the affine $XY$ plane we set $z=1$ and we obtain the components of the 
plot shown in Figure~\ref{rlwc}, that is the line $y=0$ and the curve plotted in red whose equation 
is $2x^3 + 2xy^2 + 7x^2 + 3y^2 + 8x=0$. We can also 
plot these varieties over the projective plane (using Gnomonic projection) as was shown in \cite{pjrl}, as well
as plot them over other components of the projective plane also, as the affine plane $ZY$.

\begin{figure}
\begin{center}
\includegraphics[scale=1.5]{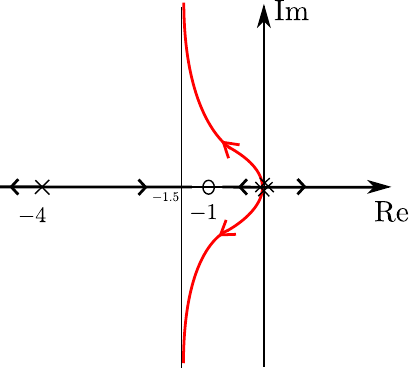} 
\caption{\label{rlwc} Root-Locus for $\displaystyle G(s)=\frac{s+1}{s^2(s+4)}$} 
\end{center}
\end{figure}

\end{example}

\section{Algebraic Dual Root-Locus -- ADRL} \label{darl}
Duality is a fundamental concept in projective algebraic geometry. In fact, it is a basic
property of the real projective plane $\mathbb{RP}^2$ that a ``point'' with (nonzero) homogeneous
coordinate $(a:b:c)$ can be associated to a ``line'' with equation $ax+by+cz=0$ in and vice-versa.
This kind of duality can be extended from a projective line to a projective plane curve $({\cal W})$ 
defined by $f(x,y,z) = 0$, where $f(x,y,z)\in\mathbb R[x,y,z]$ is a homogeneous polynomial. 
By the natural duality between lines and points in $\mathbb{RP}^2$, 
each tangent line to the curve can 
be associated to a point with homogeneous coordinate, for instance, $(u:v:w)$, and the main result is that this 
set of points is also the solution to some equation $g(u,v,w)=0$, where $g(u,v,w)\in\mathbb R[u,v,w]$ is a
homogeneous polynomial, that represents a curve (${\cal W}^*$) over $\mathbb{RP}^2$ (in fact over the dual 
of $\mathbb{RP}^2$, which it is itself). Therefore  
${\cal W}^*$ is denominated the {\em dual curve} of ${\cal W}$. Interestingly, we also have that if we take the dual of the 
dual of a curve we restore back the original curve, that is $({\cal W}^*)^* = {\cal W}$ (see \cite{gray}). Mathematically, 
the dual curve of $f(x,y,z)=0$ is the set of points $(u:v:w)=(\partial f/\partial x:\partial f/\partial y:\partial f/\partial z)$ in 
$\mathbb{PR}^2$, or $u=\lambda(\partial f/\partial x)$, $v=\lambda(\partial f/\partial y)$ and $w=\lambda(\partial f/\partial z)$,
for some $\lambda\neq 0$. To find the curve which these points belongs to, we can restate the problem as the one of 
eliminating $x,y,z$ and $\lambda$ from the set of equations 
$f(x,y,z)=0$, $u-\lambda(\partial f/\partial x)=0$, $v-\lambda(\partial f/\partial y)=0$ and 
$w-\lambda(\partial f/\partial z)=0$, which, in turn, can be solved by finding a Grobner basis for the ideal
\begin{equation}\label{dideal}
I = \left\langle f(x,y,z), u-\lambda\frac{\partial f}{\partial x}, v-\lambda\frac{\partial f}{\partial y}, 
w-\lambda\frac{\partial f}{\partial z}\right\rangle
\end{equation}

In our context, we are primarily interested in obtaining the dual curve for each irreducible component of
the RL, as computed in Section~\ref{ic} above, and collate them to construct a new RL plot that we 
denominate ``Algebraic Dual Root-Locus'' (ADRL). So, we will calculate de ideal defined in (\ref{dideal})
for the examples presented in Section~\ref{ic}.

\subsection{Examples}

\begin{example}\label{ex3}\em
Let be $G(s) = (s+1)/s^2$, whose RL plot is shown in Figure~\ref{rlc}, and
we found in Example~\ref{ex1} that the PjRL can be represented by the set
of ideals in Equations~(\ref{j1ex1},\ref{j2ex1}), that is 
\begin{eqnarray}
J_1 & = & \langle y, x^2k_d + xzk_n + z^2k_n \label{j1ex3}\rangle \\
J_2 & = & \langle x^2 + y^2 + 2xz, 2xk_d +zk_n\label{j2ex3}\rangle 
\end{eqnarray}

To find the dual curve of $\mathbb V(J_1)$, we have to calculate the dual curve of 
$f(x,y,z)=x^2k_d + xzk_n + z^2k_n$, and therefore the ideal in Equation~(\ref{dideal})
for this curve is:
\[
I = \langle x^2k_d + xzk_n + z^2k_n, u-\lambda (2xk_d+zk_n), v, w-\lambda(xk_n+2zk_n)\rangle
\]
Finding a Grobner basis for this ideal with ``Lex" monomial ordering with 
$x>y>z>k_d>k_n>\lambda>u>v>w$ we eliminate $x,y, z, \lambda$ and obtain the curve:
\[
g(u,v,w) = k_dw^2+k_nu^2-k_nuw
\]
So we can define the ``dual ideal" for $J_1$ as:
\begin{equation}\label{j1dex3}
J_{1}^{d} = \langle v, k_dw^2+k_nu^2-k_nuw\rangle
\end{equation}
and the dual curve for $\mathbb V(J_1)$ is $\mathbb V(J_{1}^{d})$.

To find the dual curve of $\mathbb V(J_2)$, we have to calculate the dual curve of 
$f(x,y,z)=x^2 + y^2 +2xz$, and therefore the ideal in Equation~(\ref{dideal})
for this curve is:
\[
I = \langle x^2 + y^2 + 2xz, u-\lambda (2x + 2z), v - \lambda(2y), w-\lambda(2x)\rangle
\]
Finding a Grobner basis for this ideal, to eliminate $x,y,z,\lambda$, we obtain
\[
h(u,v,w) = v^2+2uw-w^2
\]
To find the parametrization for $h$ above we use the ideal
as shown below
\[
I = \langle 2xk_d +zk_n, u-\lambda (2x + 2z), v - \lambda(2y), w-\lambda(2x)\rangle
\]
and finding a Grobner basis for this ideal, to eliminate $x,y,z,\lambda$, we obtain
the (parametrized) curve $h_1(u,v,w) = k_nu + (2k_d - k_n)w$  and then we can define 
the again ``dual ideal" for $J_2$ as:
\begin{equation}\label{j2dex3}
J_{2}^{d} = \langle v^2+2uw-w^2, k_nu + (2k_d - k_n)w\rangle
\end{equation}
Now we can define the Algebraic Dual Root-Locus for $G(s)=(s+1)/s^2$ as the union 
of the varieties $\mathbb V(J_1^d)$ and $\mathbb V(J_2^d)$ where $J_1^d$ and $J_2^d$
are defined in (\ref{j1dex3}) and (\ref{j2dex3}), respectively. Below we analyse each variety 
in order to obtain the complete plot for the ADRL:

\begin{description}
\item[Variety $\mathbb V(J_1^d)$:] Defined by the ideal $J_1^d$, as shown in Equation~(\ref{j1dex3})
\begin{itemize}
\item Initial Points: $k_d=1$ and $k_n=0$. From (\ref{j1dex3}), we get $v=0$ and $w^2=0$ or $w=0$. 
Therefore the initial point 
for $\mathbb V(J_1)$ is $(1:0:0)$, a point at infinity, or the intersection of horizontal lines in affine $UV$ plane.
\item Terminal points: $k_d=0$ and $k_n=1$. We get from (\ref{j1dex3}), $v=0$ and $u^2 - uw=0$. Then we have (a) $u=0$ 
and $w=1$, which is
the point $(0:0:1)$, or the point $(0,0)$ in affine plane $UV$ and (b) $u=w=1$ which is the point $(1:0:1)$ or $(1,0)$ 
in affine $UV$ plane.
\item Intermediary Points: $k_d=1$ and $k_n=\lambda\neq 0$. Again from (\ref{j1dex3}), we have $v=0$ and 
$\lambda u^2 - \lambda uw + w^2=0$,
and we must have $w=1$ ($w=0$ would imply $u=0$ what is impossible); so, all intermediary points are at finite position 
($w=1$) and is given by
$v=0$ and $u^2 - u +1/ \lambda=0$, $\lambda\neq 0$. Then $u=\frac{1\pm\sqrt{1-4/\lambda}}{2}$, which give us the 
ADRL over $v=0$ line in affine $UV$ plane.

\end{itemize}

\item[Variety $\mathbb V(J_2^d)$:] Defined by the ideal $J_2^d$, as shown in Equation~(\ref{j2dex3})
\begin{itemize}
\item Initial Points: $k_d=1$ and $k_n=0$. From (\ref{j2dex3}), we get $v^2 + 2uw - w^2=0$ and $2w=0$ or $w=0$. Then we 
have $v^2=0$
or $v=0$. Therefore the initial point for $\mathbb V(J_2)$ is $(1:0:0)$, a point at infinity (intersection of horizontal 
lines) in $UV$ plane.
\item Terminal points: $k_d=0$ and $k_n=1$. We get from (\ref{j2dex3}), $v^2 + 2uw - w^2=0$ and $u-w=0$ or 
$u=w$. Then we get $v^2 + u^2=0$ what implies $u=v=w=0$ which is not allowed, so this variety has no terminal points.
\item Intermediary Points: $k_d=1$ and $k_n=\lambda\neq 0$. Again from (\ref{j2dex3}), we have $v^2 + 2uw - w^2=0$ and 
$\lambda u +(2 - \lambda)w=0$, and we must have $w=1$ ($w=0$ would imply $u=v=0$ what is impossible); so, all intermediary 
points are at finite position and are given by $v^2 + 2u -1 = 0$ and $u = 1 - 2/\lambda$, $\lambda\neq 0$, which is the 
parametrized equation of the parabola as shown in ADRL plot.

\end{itemize}

\end{description}

Now, since the complete ADRL is $\mathbb V(J_1^d) \cup \mathbb V(J_2^d)$, we have:
\begin{itemize}
\item Initial points: $(1:0:0)$ (from $\mathbb V(J_1^d)$) plus $(1:0:0)$ (from $\mathbb V(J_2^d)$); so we have a duplicate 
point at $(1:0:0)$ or at infinity in affine plane $UV$.
\item Terminal Points: $\{(0:0:1), (1:0:1)\}$ only from $\mathbb V(J_1^d)$; or $(0.0)$ and $(1,0)$ in $UV$ plane 
\item Intermediary points: $\{(u:0:1), u=(1\pm\sqrt{1-4/\lambda})/2, \lambda\neq 0\}$ from 
$\mathbb V(J_1^d)$ plus $\{(u:v:1), u= 1 -2/\lambda, \lambda\neq 0$ and $v^2+2u-1=0$\} from $\mathbb V(J_2^d)$, in
$UV$ plane.
\end{itemize}
The ADRL plot for $G(s) = (s+1)/s^2$ is shown in Figure~\ref{darlex3}. 

It is important to note that if we take the duals of $J_1^d$ and $J_2^2$, defined in Equations (\ref{j1dex3}) and 
(\ref{j2dex3}), respectively, we get back, the ideals $J_1$ and $J_2$, as defined in Equations (\ref{j1ex3}) and 
(\ref{j2ex3}), respectively. The procedure for doing that is eliminating $u,v,w$ and $\lambda$ in the ideals $I_1$, 
$I_2$ and $I_3$ (defined below), using Lex monomial ordering with $u>v>w>x>y>z>k_d>k_n>\lambda$.
\begin{eqnarray*}
I_1 & = & \langle  k_dw^2+k_nu^2-k_nuw, x-\lambda(2uk_n-k_nw), y, z-\lambda(2k_dw-k_nu)\rangle\\
I_2 & = & \langle  v^2+2uw-w^2, x-\lambda(2w), y-\lambda(2v), z-\lambda(2u-2w)\rangle \\
I_3 & = & \langle  k_nu + (2k_d - k_n)w, x-\lambda(2w)y-\lambda(2v), z-\lambda(2u-2w)\rangle
\end{eqnarray*}

\begin{figure}
\begin{center}
\includegraphics[scale=1.5]{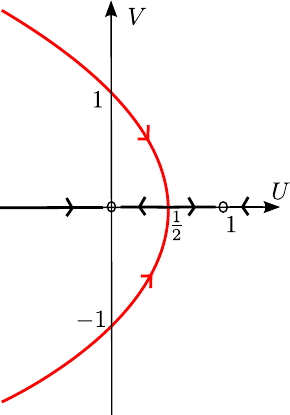} 
\caption{\label{darlex3} Algebraic Dual Root-Locus for $G(s)=(s+1)/s^2$} 
\end{center}
\end{figure}

\end{example}

\begin{example} \em \label{ex4} We now consider $\displaystyle G(s)=\frac{s+1}{s^2(s+4)}$, 
whose RL plot is shown in Figure~\ref{rlwc} and whose PjRL is represented by the ideals 
in Equations~(\ref{j1ex2},\ref{j2ex2}):
\begin{eqnarray*}
J_1 & = & \langle y, x^3k_d + 4x^2zk_d + xz^2k_n + z^3k_n\rangle \\
J_2 & = & \langle 2x^3 + 2xy^2 + 7x^2z + 3y^2z + 8xz^2, 3x^2k_d - y^2k_d + 8xzk_d + z^2k_n\rangle
\end{eqnarray*}

Repeating the reasoning used in Example~\ref{ex3} above, we obtain the ``duals'' ideals:
\begin{eqnarray}
J_1^d & = & \langle v,4k_duw^2 - k_dw^3 + k_nu^3- k_nu^2w\rangle \label{j1dex4}\\
J_2^d & = & \langle f(u,v,w), g(u,v,w)\rangle \label{j2dex4}
\end{eqnarray}
where $f$ and $g$ are the ``astonishing'' polynomials\footnote{In fact, a well known result in projective algebraic geometry
states that if a curve has degree $d$ (and no singularities) its dual has degree $d(d-1)$ \cite[pp.~173]{gray}; 
in our case the polynomial in the ideal $J_2$ has degree $3$, so its dual has degree $6$.}
\begin{eqnarray*}
f(u,v,w) & = & 216u^5w+144u^4v^2-621u^4w^2-912u^3v^2w+720u^3w^3-352u^2v^4+718u^2v^2w^2 - \\
         &   & 424u^2w^4 -232uv^4w-176uv^2w^3+128uw^5-240v^6+107v^4w^2-8v^2w^4-16w^6 \\
g(u,v,w) & = & 294912k_d^4u^3w-327680k_d^4u^2v^2-798720k_d^4u^2w^2+1671168k_d^4uv^2w+512000k_d^4uw^3-\\
              &    & 1048576k_d^4v^4-675840k_d^4v^2w^2-96000k_d^4w^4+73728k_d^3k_nu^4-448512k_d^3k_nu^3w+\\
              &    & 430080k_d^3k_nu^2v^2+706304k_d^3k_nu^2w^2-788480k_d^3k_nuv^2w-403200k_d^3k_nuw^3+ \\
              &    & 131072k_d^3k_nv^4+300288k_d^3k_nv^2w^2+75600k_d^3k_nw^4-46656k_d^2k_n^2u^4+174816k_d^2k_n^2u^3w-\\
              &    & 68928k_d^2k_n^2u^2v^2-224292k_d^2k_n^2u^2w^2+109920k_d^2k_n^2uv^2w+119040k_d^2k_n^2uw^3- \\
              &    & 6144k_d^2k_n^2v^4-41364k_d^2k_n^2v^2w^2-22320k_d^2k_n^2w^4+8208k_dk_n^3u^4-26172k_dk_n^3u^3w+\\ 
              &    & 3088k_dk_n^3u^2v^2+30636k_dk_n^3u^2w^2-4764k_dk_n^3uv^2w-15616k_dk_n^3uw^3+ \\
              &    & 128k_dk_n^3v^4+1788k_dk_n^3v^2w^2+2928k_dk_n^3w^4-441k_n^4u^4+1344k_n^4u^3w- \\
              &    & 42k_n^4u^2v^2-1528k_n^4u^2w^2+64k_n^4uv^2w+768k_n^4uw^3-k_n^4v^4-24k_n^4v^2w^2-144k_n^4w^4
\end{eqnarray*}
Remembering that the dual curve for $2x^3 + 2xy^2 + 7x^2z + 3y^2z + 8xz^2=0$ is $f(u,v,w)=0$ and
that $g(u,v,w)=0$ is used just for obtaining a parametrization. The plot for the correspondig ADRL, 
obtained by the Scilab package (\cite{scilab}) is shown in Figure~\ref{darlex4}.

\begin{figure}
\begin{center}
\includegraphics[scale=1]{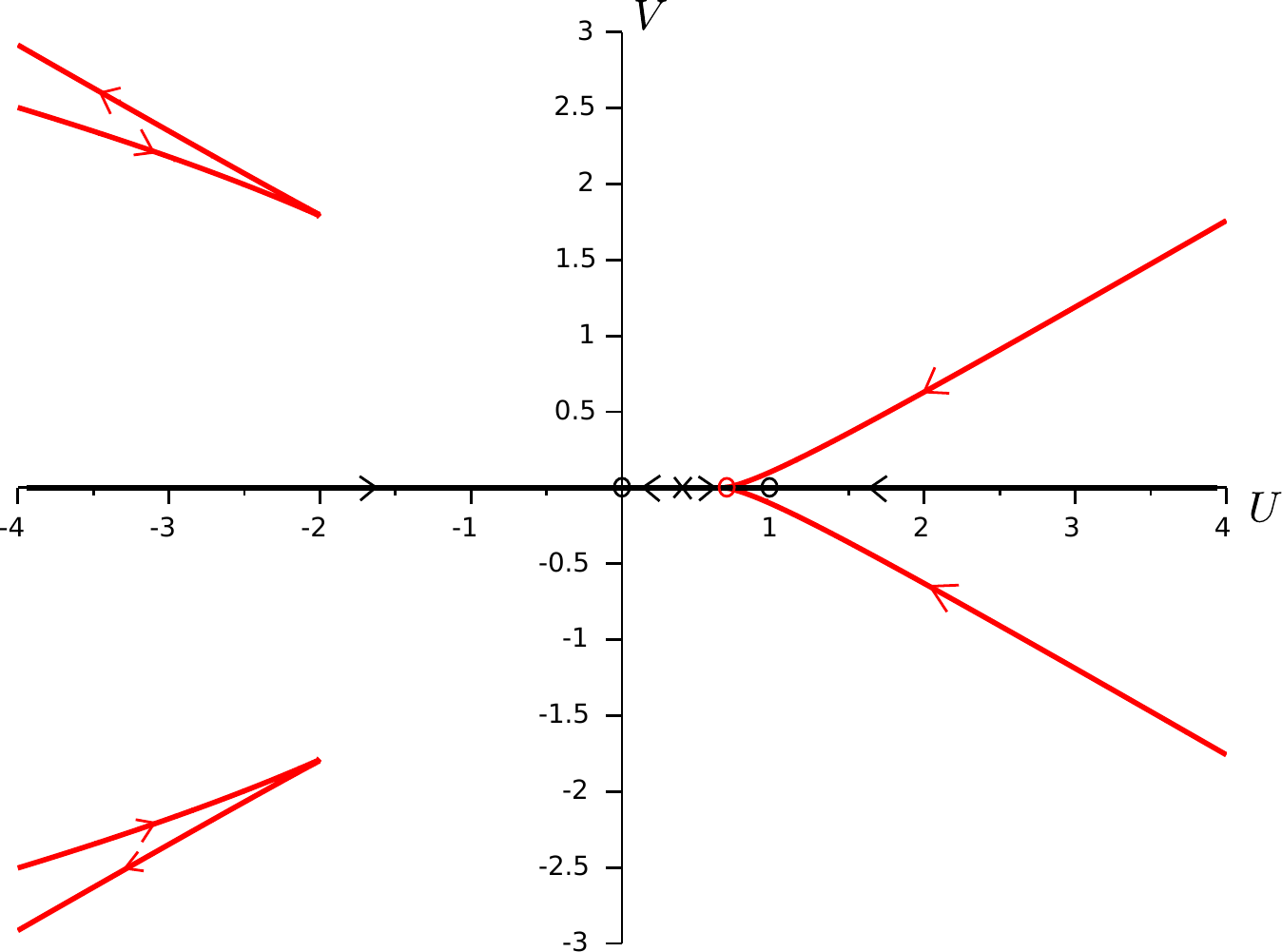}
\caption{\label{darlex4} Algebraic Dual Root-Locus for $\displaystyle G(s)=\frac{s+1}{s^2(s+4)}$} 
\end{center}
\end{figure} 

\end{example}

\section{Conclusions}
We have presented in this paper a procedure for isolating the planes curves that makes up 
the root-locus plot for an irreducible transfer function. This procedure can be easily implemented 
in a computational algebra software package. We also showed how to compute the dual curve
(in projective algebraic geometry sense) to each plane curve and join them to compose what 
we denominated ``Algebraic Dual Root Locus'' (ADRL). Some examples were worked out in order to show 
the effectiveness of the procedure. We intend to investigatethe properties of the ADRL 
more deeply in future works.

\bibliographystyle{plain}

\end{document}